\documentclass{moriond}

\usepackage[utf8]{inputenc}
\usepackage[english]{babel}    
\usepackage{amssymb,amsmath} 
\usepackage{bbold}

\def\Journal#1#2#3#4{{#1} {\bf #2}, #3 (#4)}

\def\NPB{{\em Nucl. Phys.} B}
\def\PLB{{\em Phys. Lett.}  B}
\def\PRL{\em Phys. Rev. Lett.}
\def\PRD{{\em Phys. Rev.} D}

\def\be{\begin{equation}}
\def\ee{\end{equation}}
\def\bea{\begin{eqnarray}}
\def\eea{\end{eqnarray}}

\newcommand{\Photo}{}

\begin{document}
\begin{flushright}
 IFT-UAM/CSIC-14-049\\
 FTUAM-14-20
\end{flushright}

\vspace*{4cm}
\title{LEPTON FLAVOUR VIOLATING HIGGS DECAYS}

\author{ E. ARGANDA$^{1}$, M.J. HERRERO$^{2}$, X. MARCANO$^{2}$, C. WEILAND$^{2}$\footnote{Talk given by C. Weiland at the 49th Rencontres de Moriond on Electroweak Interactions and Unified Theories.} }

\address{$^1$Departamento de F\'{\i}sica Te\'orica, Facultad de Ciencias,\\
Universidad de Zaragoza, 50009 Zaragoza, Spain\\
$^2$Departamento de F\'{\i}sica Te\'orica and Instituto de F\'{\i}sica Te\'orica, IFT-UAM/CSIC,\\ Universidad Aut\'onoma de Madrid, Cantoblanco, 28049 Madrid, Spain}

\maketitle
\abstracts{We present our study of lepton flavour violating decays of a Higgs boson with properties compatible with those of the particle recently discovered at the LHC. We worked in the context of the inverse seesaw model, considering the most generic case where the Standard Model is extended by three pairs of fermionic singlets in order to generate the neutrino masses and mixings required by neutrino oscillations. Using a full one-loop calculation together with the most recent experimental and theoretical constraints, we discuss the dependence on the parameters of the inverse seesaw model before concluding on the largest allowed branching ratios through scans over the full parameter space.}

\section{Introduction}

During the past twenty years, neutrino experiments have harvested a vast number of exciting results and, nowadays, neutrino oscillations are a well studied phenomena whose
parameters have all been precisely measured, with the exception of a CP violating phase~\cite{GonzalezGarcia:2012sz}. This corresponds to the indisputable observation of lepton
flavour violation (LFV) in the neutral sector. We are then forced to ask if charged lepton flavour could be violated too. Indeed, once non-zero masses and mixings in the
neutrino sector are taken into account, charged LFV (cLFV) can arise at the one-loop level. However, in the Standard Model (SM), these signals are strongly suppressed by a GIM
mechanism, making them unobservable at any current or planned experiment. Thus, the detection of a cLFV process would provide a clear evidence of new Physics. This has motivated numerous experiments in the past~\cite{Bernstein:2013hba}, most of them focusing on radiative or three-body lepton decays or neutrinoless $\mu - e$ conversion in muonic atoms.

Two years ago, the CMS and ATLAS experiments announced the discovery of a new particle and they have started a vast experimental program aimed at measuring its properties. Their
latest results point toward a mass between $m_h=125.5\pm 0.6 $GeV (ATLAS~\cite{Aad:2013wqa}) and $m_h=125.7\pm 0.4 $GeV (CMS~\cite{Chatrchyan:2013lba}) and characteristics otherwise compatible with the Higgs boson of the Standard Model. Both experiments are actively searching for leptonic decays of the Higgs boson and they have evidence for its decay into $\tau$ leptons. This makes the search for cLFV Higgs decays very timely and complementary to other cLFV searches.

The Standard Model cannot accommodate neutrino masses, which are needed to explain neutrino oscillations. This calls for new Physics in the leptonic
sector and one of the simplest extensions of the SM is 
the addition of right-handed (RH) neutrinos, which are fermionic gauge singlets. In this work, we consider the Inverse Seesaw (ISS) 
model~\cite{Mohapatra:1986aw,Mohapatra:1986bd,Bernabeu:1987gr} whose main advantage over the usual type I seesaw is that it naturally allows for large neutrino Yukawa couplings, of order $\mathcal{O}(1)$, and a seesaw scale close to the electroweak scale. This makes the ISS simultaneously testable at the LHC, through the direct production of the RH neutrinos, and at low-energy experiments via loop-generated effects. 

In our study, we have considered the complete set of one-loop diagrams contributing to cLFV Higgs decays which can be found with the corresponding formulas in our main
article~\cite{Arganda:2014dta}, together with our full numerical results.

\section{The inverse seesaw model}

Neutrino oscillations are clearly established nowadays and can easily be explained by the existence of massive neutrinos with intergenerational mixing. However, neutrino masses cannot be generated in the SM due to the absence of fermionic singlets or $\mathrm{SU}(2)$ triplets. The inverse seesaw addresses this shortcoming of the SM by adding two types of fermionic singlets, $\nu_R$ and $X$ with opposite lepton numbers, per generation. The corresponding Lagrangian is
\begin{equation}
 \mathcal{L}_\mathrm{ISS}=-Y^{ij}_\nu\overline{L_{i}}\widetilde{\Phi}\nu_{Rj}-M_R^{ij}\overline{\nu_{Ri}^C} X_j-\frac{1}{2}\mu_{X}^{ij}\overline{X_{i}^C} X_{j}+ h.c.\,,
\end{equation}
where $L$ is the SM lepton doublet, $\Phi$ is the SM Higgs doublet, $\widetilde{\Phi}=\imath \sigma_2 \Phi^*$, with $\sigma_2$ the corresponding Pauli matrix, $Y_\nu$ is the  neutrino Yukawa coupling matrix, $M_R$ is a lepton number conserving mass matrix, and $\mu_X$ is a Majorana mass matrix that violates lepton number conservation by two units. This leads to the following neutrino mass matrix in the $(\nu_L^C\,,\;\nu_R\,,\;X)$ basis, after electroweak symmetry breaking,
\begin{equation}
 M_{\mathrm{ISS}}=\left(\begin{array}{c c c} 0 & m_D & 0 \\ m_D^T & 0 & M_R \\ 0 & M_R^T & \mu_X \end{array}\right)\,,
\end{equation}
with $m_D=Y_\nu \langle \Phi\rangle$, where the Higgs vacuum expectation value is taken to be $\langle \Phi\rangle=v = 174\,\mathrm{GeV}$. Considering only one generation and making the natural assumption $\mu_X \ll m_D, M_R$, the diagonalization of the mass matrix gives the following mass eigenstates
\begin{align}
 m_\nu&=\frac{m_{D}^2}{m_{D}^2+M_{R}^2} \mu_X\,,\\
 m_{N_1,N_2}&=\pm \sqrt{M_{R}^2+m_{D}^2} + \frac{M_{R}^2 \mu_X}{2 (m_{D}^2+M_{R}^2)}\,.
\end{align}
What makes the ISS mechanism attractive is the fact that the smallness of the light neutrino mass is directly proportional to the smallness of $\mu_X$, the parameter that controls the
size of the lepton number violating mass term. As such, its smallness is natural in the sense of 't~Hooft~\cite{'tHooft:1979bh}. Moreover, the presence of this extra parameter
decouples the Weinberg operator which generates the light neutrino masses from the higher dimensional operators responsible for low-energy effects like cLFV and lepton
universality violation. Thus, it is natural to expect that lepton flavour violating Higgs decays will be strongly enhanced in the inverse seesaw model.

\section{Lepton flavour violating Higgs decays in the inverse seesaw}

We have implemented the complete set of one-loop diagrams in our private \texttt{Mathematica} code, considering a Higgs boson with a mass $m_H=126\,\mathrm{GeV}$ whose total SM
decay width was computed using \texttt{FeynHiggs}~\cite{Heinemeyer:1998yj,Heinemeyer:1998np,Degrassi:2002fi}. We have also included the relevant experimental constraints,
starting with neutrino oscillation data as given by the NuFit collaboration in the v1.2 of their results~\cite{GonzalezGarcia:2012sz} and the upper limit on the effective
electron neutrino mass in $\beta$ decays from the Mainz and Troitsk experiment~\cite{Kraus:2004zw,Aseev:2011dq}. This was done by using a modified Casas-Ibarra parametrization
as described in our article~\cite{Arganda:2014dta}, whose validity was checked by requiring that the difference between the input and output light neutrino masses was below $10\%$
and that the full $9\times 9$ rotation matrix was unitary. Since arbitrary large Yukawa couplings can be generated when using the Casas-Ibarra parametrization, we have
required, for $i,j=1,2,3$,
\begin{equation}
\frac{|Y_{ij}|^2}{4\pi}<1.5\,,
\end{equation}
ensuring that the neutrino Yukawa couplings do not leave the perturbative regime.

In addition to the above mentioned requirements, we have also implemented constraints coming from the LHC and low-energy experiments. First, RH neutrinos lighter than the
Higgs boson could open new invisible decay channels, strongly enhancing the Higgs invisible decay width in some case. In order to avoid this, we
have required that sterile neutrinos are heavier than $200\,\mathrm{GeV}$, thus escaping these potential constraints. Having simultaneously large neutrino Yukawa couplings and
RH neutrinos with a mass close to the electroweak scale would generate large cLFV branching ratios. We have implemented the one-loop computation of the
$\ell_i \rightarrow \ell_j \gamma$ decay rates within the same framework using standard analytical formulas~\cite{Ilakovac:1994kj,Deppisch:2004fa} and have applied
the upper bounds on cLFV radiative
decays coming from the MEG~\cite{Adam:2013mnn} (${\rm BR}(\mu\to e\gamma)\leq 5.7\times 10^{-13}$ at $90\%$ CL) and BaBar~\cite{Aubert:2009ag} experiments 
(${\rm Br}(\tau \to \mu \gamma)<4.4 \times 10^{-8}$, ${\rm Br}(\tau \to e \gamma)<3.3 \times 10^{-8}$ at $90\%$ CL).
If large contributions to
cLFV processes can be generated, a large contribution to lepton EDMs could also be expected in the general case. To avoid this, we assume in most of our study that all mass
matrices and the PMNS matrix are real. It was recently shown that lepton universality tests provide constraints complementary to the one derived from cLFV
processes~\cite{Abada:2012mc,Abada:2013aba}. However, since we consider only RH neutrinos heavier than the Higgs boson, points in the parameter space that are excluded by
lepton universality test are also excluded by radiative cLFV decays. In the end, we found that the most constraining observable for our study is $\mu \rightarrow e \gamma$, due
to the stringent upper limit obtained by the MEG collaboration.

We have focused on the decays $H\rightarrow \mu \bar \tau,e\bar\tau,e\bar\mu$. While this is not necessarily the case in the most general scenario, their branching
ratios are equal to the ones of their CP conjugates under our assumptions of real PMNS and mass matrices. We have distinguished two cases in our study: degenerate or hierarchical
heavy neutrinos. We will present here our results for the degenerate scenario, since they illustrate well the main features of our full study. A discussion of cLFV Higgs in both degenerate and hierarchical cases can be found in our main article~\cite{Arganda:2014dta}.

\begin{figure}[t!] 
\begin{center}
\begin{tabular}{cc}
\includegraphics[width=0.474\textwidth]{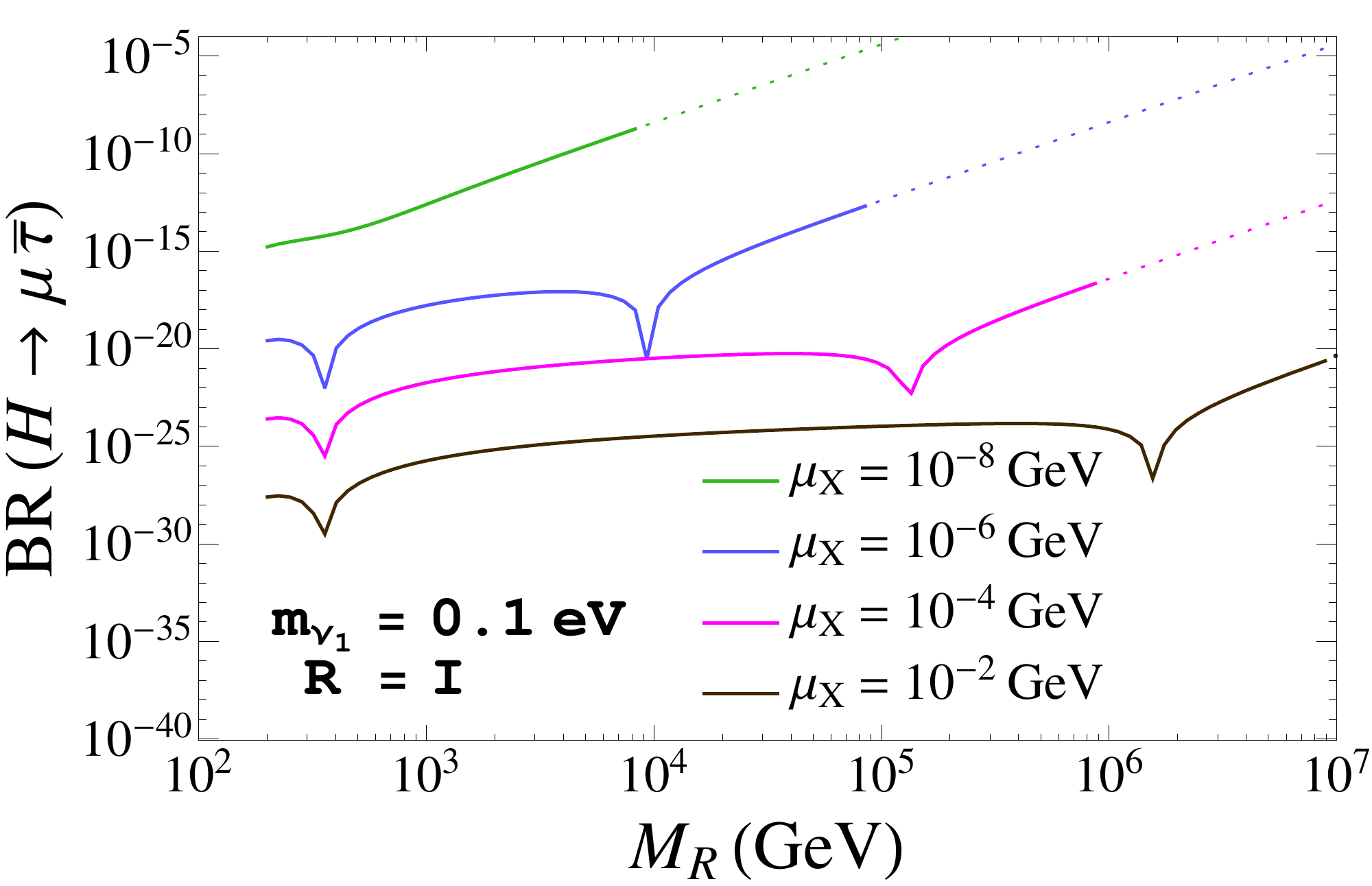} &
\includegraphics[width=0.474\textwidth]{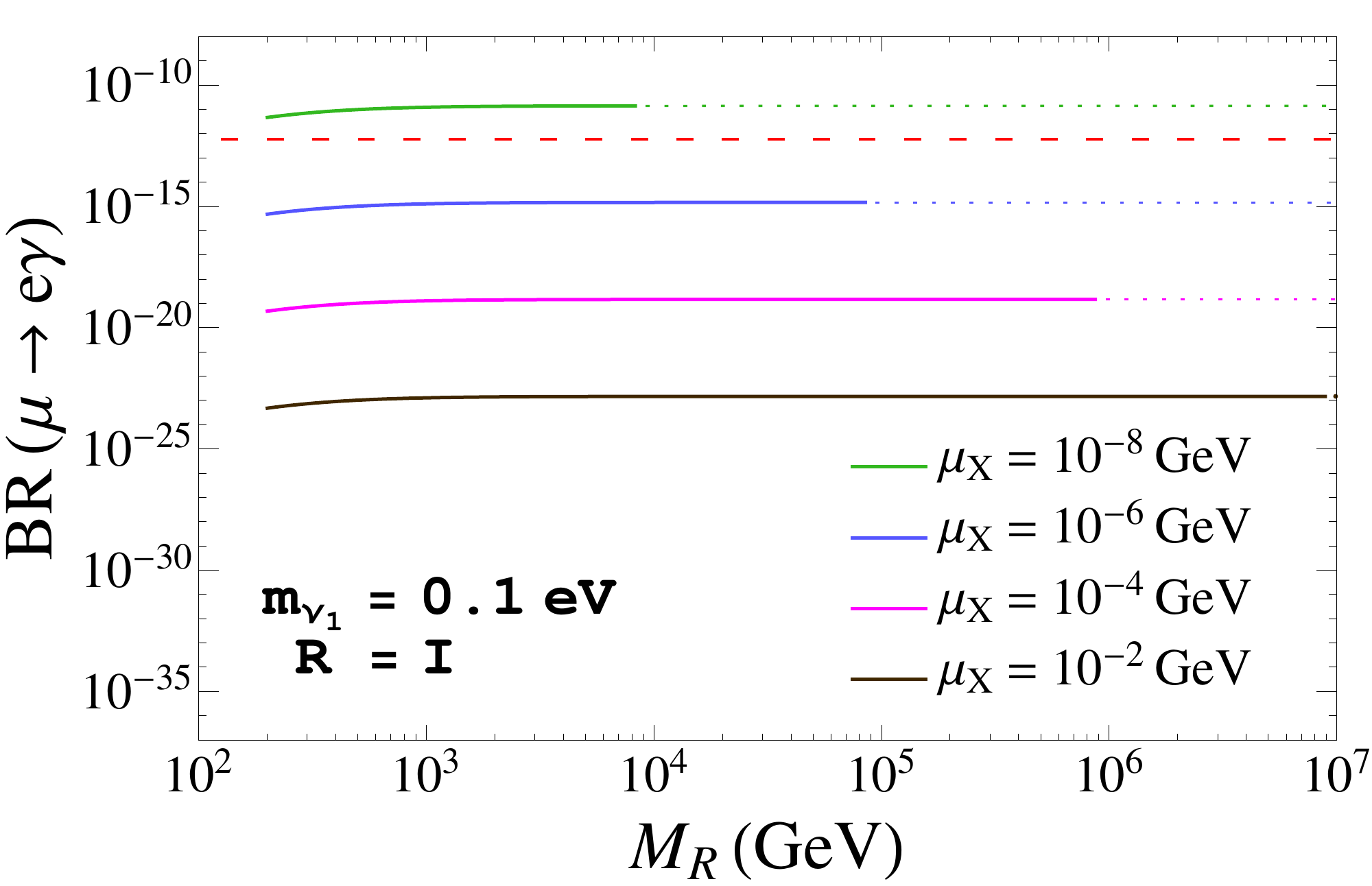}
\end{tabular}
\caption{Branching ratios of $H\rightarrow\mu\bar\tau$ and $\mu\rightarrow e \gamma$ as a function of $M_R$ for different values of $\mu_X$. Dotted lines correspond to non-perturbative neutrino Yukawa couplings while the red dashed line is the experimental upper bound from MEG. }\label{BRs_MR_degenerate}
\end{center}
\end{figure}
Let us start by studying the dependence of cLFV decays on the parameters of the ISS, as can be seen in Fig.~\ref{BRs_MR_degenerate} where we have plotted the decays $H\rightarrow \mu \bar \tau$, the cLFV Higgs decay with the largest branching ratio, and $\mu \rightarrow e \gamma$, the most constraining radiative cLFV observable, in a degenerate scenario. Degenerate heavy neutrinos were obtained by choosing degenerate entries in $M_R={\rm diag}(M_{R_1},M_{R_2},M_{R_3})$ and in $\mu_X={\rm diag}(\mu_{X_1},\mu_{X_2},\mu_{X_3})$, i.e., by setting $M_{R_i} = M_R$ and $\mu_{X_i} = \mu_X$ 
($i = 1, 2, 3$). First, we can see that smaller values of $\mu_X$ correspond to larger cLFV branching ratios. In
Fig.~\ref{BRs_MR_degenerate}, $m_{\nu_1}$ is fixed to $0.1\,\mathrm{eV}$, thus decreasing $\mu_X$ will increase the size of the neutrino Yukawa couplings $Y_\nu$, leading in
turn to larger cLFV decay rates. Second, cLFV Higgs and radiative decays present qualitatively different behaviours as functions of $M_R$. $\mathrm{BR}(\mu \rightarrow e \gamma)$ exhibits a very mild dependence on $M_R$, being constant for values $M_R \geq 10^3\,\mathrm{GeV}$. As a consequence, $\mu \rightarrow e \gamma$ will mostly constrain $\mu_X$.
However, this should not be interpreted as a non-decoupling behaviour but it is an artefact originating from the use of a modified Casas-Ibarra parametrization where keeping
$m_{\nu_1}$ fixed will lead to an increase in the neutrino Yukawa couplings when $M_R$ increases. We have explicitly checked that at large $M_R$, the cLFV radiative decays
exhibit the expected behaviour, going like
\begin{equation}
{\rm BR}^{\rm approx}_{\ell_m \rightarrow \ell_k \gamma}=8\times 10^{-17} \frac{m_{\ell_m}^5({\rm GeV}^5)}{\Gamma_{\ell_m}{\rm (GeV)}} 
\bigg|\frac{v^2}{2M_R^2}(Y_\nu Y_\nu^\dagger)_{km}\bigg|^2.
\label{approxformula} 
\end{equation}
On the contrary, $\mathrm{BR}(H\rightarrow \mu \bar \tau)$ exhibits a distinct behaviour, with a different dependence on the model parameters for various values of
$M_R$ . At large $M_R$, it grows as $M_R^4$, reaching its maximal value when the neutrino Yukawa couplings reach the perturbativity limit. There are also dips that we have
identified as coming from an interference between the dominating diagrams when $M_R$ is large and from an interference between the other diagrams when $M_R\sim 300\,\mathrm{GeV}$.

\begin{figure}[t!] 
\begin{center}
\includegraphics[width=.6\textwidth]{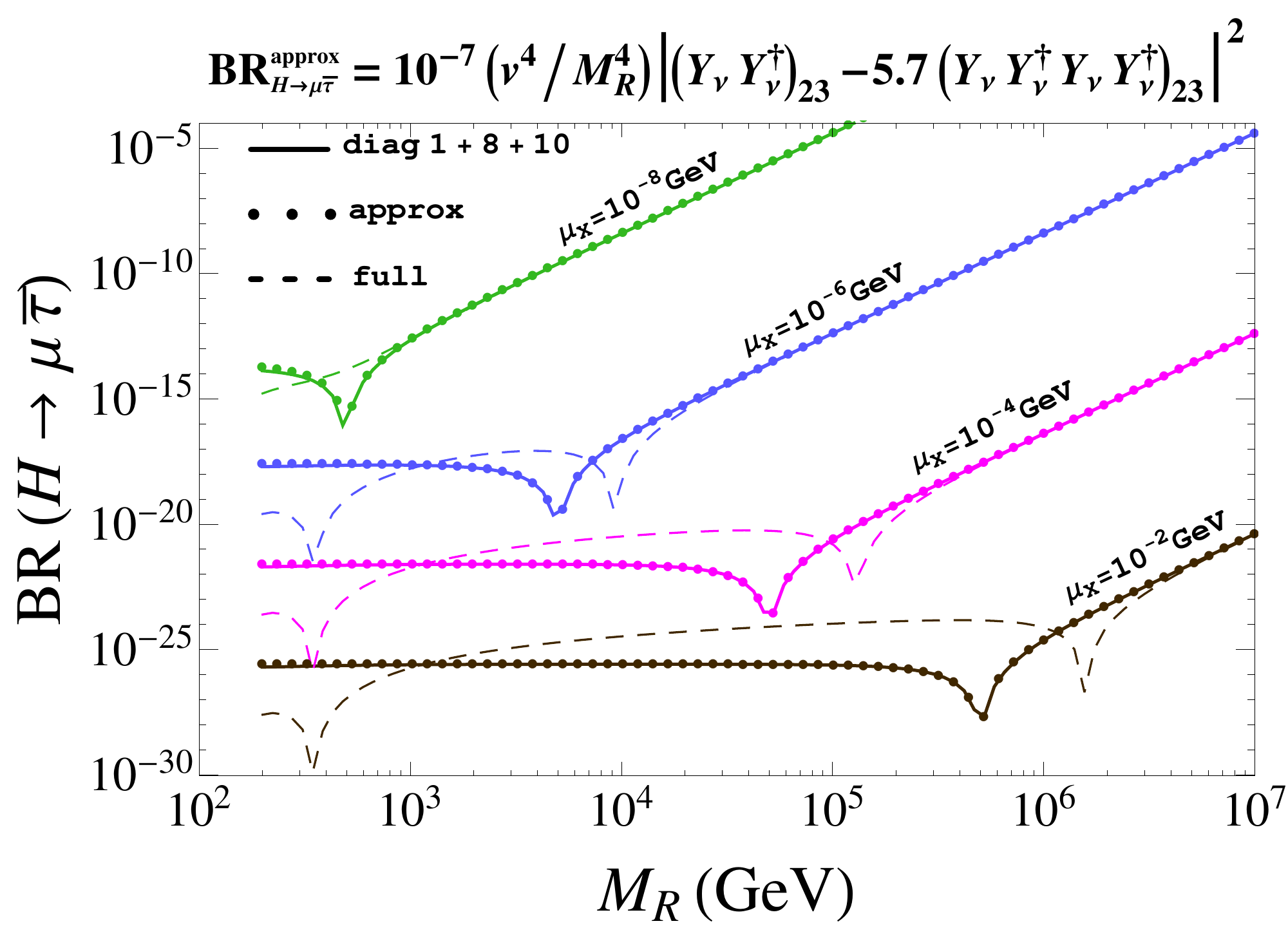}
\caption{Comparison between the full calculation (dashed line), the dominant contribution at large $M_R$ (full line) and our approximate formula (dotted line) for $\mathrm{BR}(H\rightarrow \mu \bar \tau)$ as a function of $M_R$.}
\label{fitLFVHD}
\end{center}
\end{figure}

The observed functional behaviour of $\mathrm{BR}(H\rightarrow \mu \bar \tau)$ implies that it is not simply proportional to $|Y_\nu Y_\nu^\dagger/M_R^2|^2$ as the radiative decays are. We have further explored this by focusing on the diagrams that dominate at large $M_R$ and have isolated their contribution in Fig.~\ref{fitLFVHD}
where it is compared with the full calculation and with the following approximate formula
\begin{equation}\label{FIThtaumu}
{\rm BR}^{\rm approx}_{H\to\mu\bar\tau}=10^{-7}\frac{v^4}{M_R^4}~\Big|(Y_\nu Y_\nu^\dagger)_{23}-5.7(Y_\nu Y_\nu^\dagger Y_\nu Y_\nu^\dagger)_{23}\Big|^2\,,
\end{equation}
which reproduces extremely well the dominant contribution at large $M_R$. This approximate formula can be understood by using the mass insertion approximation (MIA). Indeed, at the lowest order in the MIA, this contribution takes a form similar to the dimension 6 operator governing radiative cLFV decays, which gives the first term of Eq.~\ref{FIThtaumu}. But there are also higher order contributions like the one corresponding to diagrams with two chirality flipping mass insertions on the internal neutrino line of a loop leading to the second term in Eq.~\ref{FIThtaumu}. Besides, if the two contributions have opposite signs they will interfere destructively, leading to dips that verify $M_R^{-2} \mu_X \sim \mathrm{constant}$, which explains the dips at large $M_R$ in Figs.~\ref{BRs_MR_degenerate} and~\ref{fitLFVHD}.

\begin{figure}[t!] 
\begin{center}
\begin{tabular}{cc}\includegraphics[width=.45\textwidth]{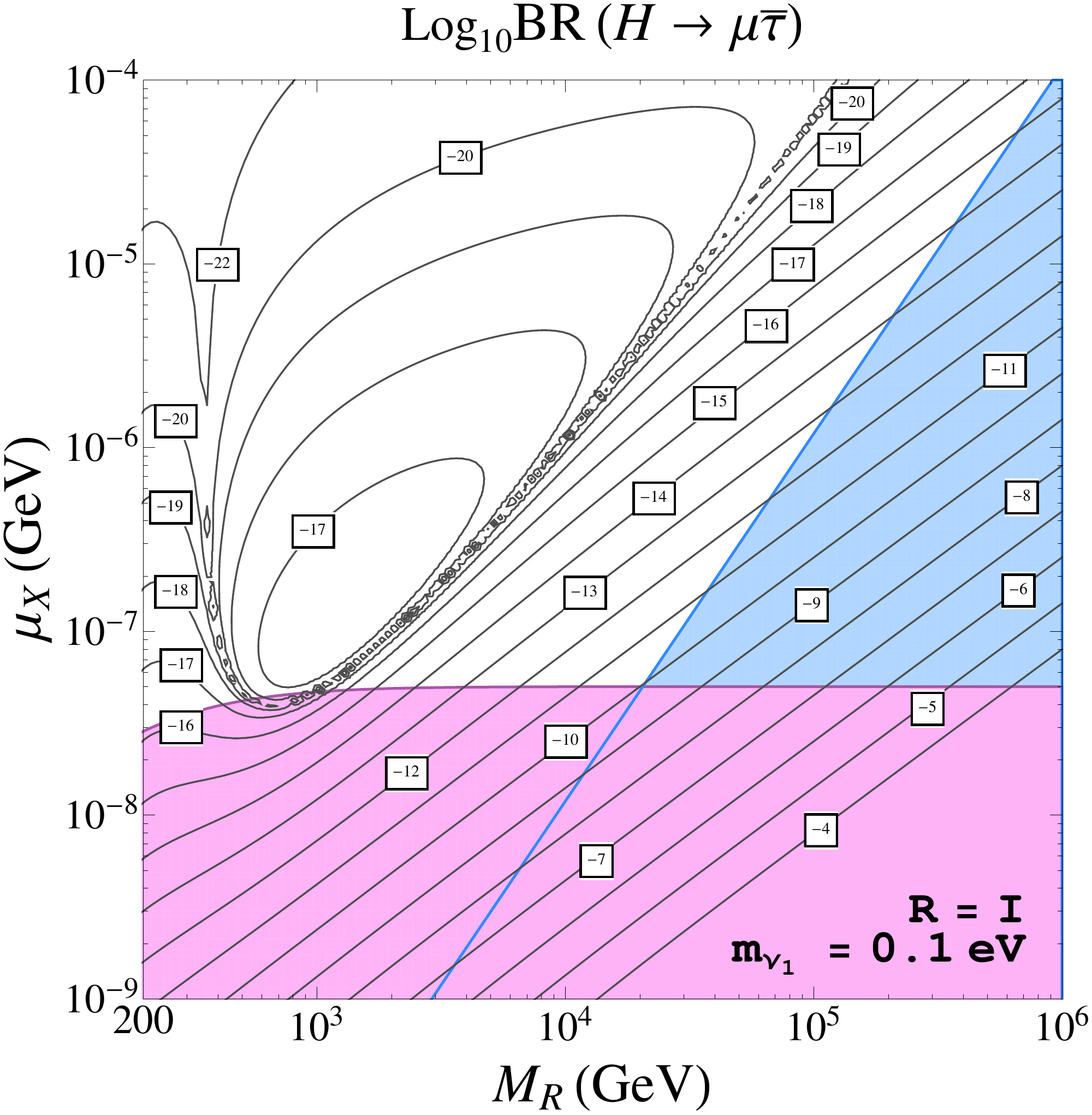} &
\includegraphics[width=.49\textwidth]{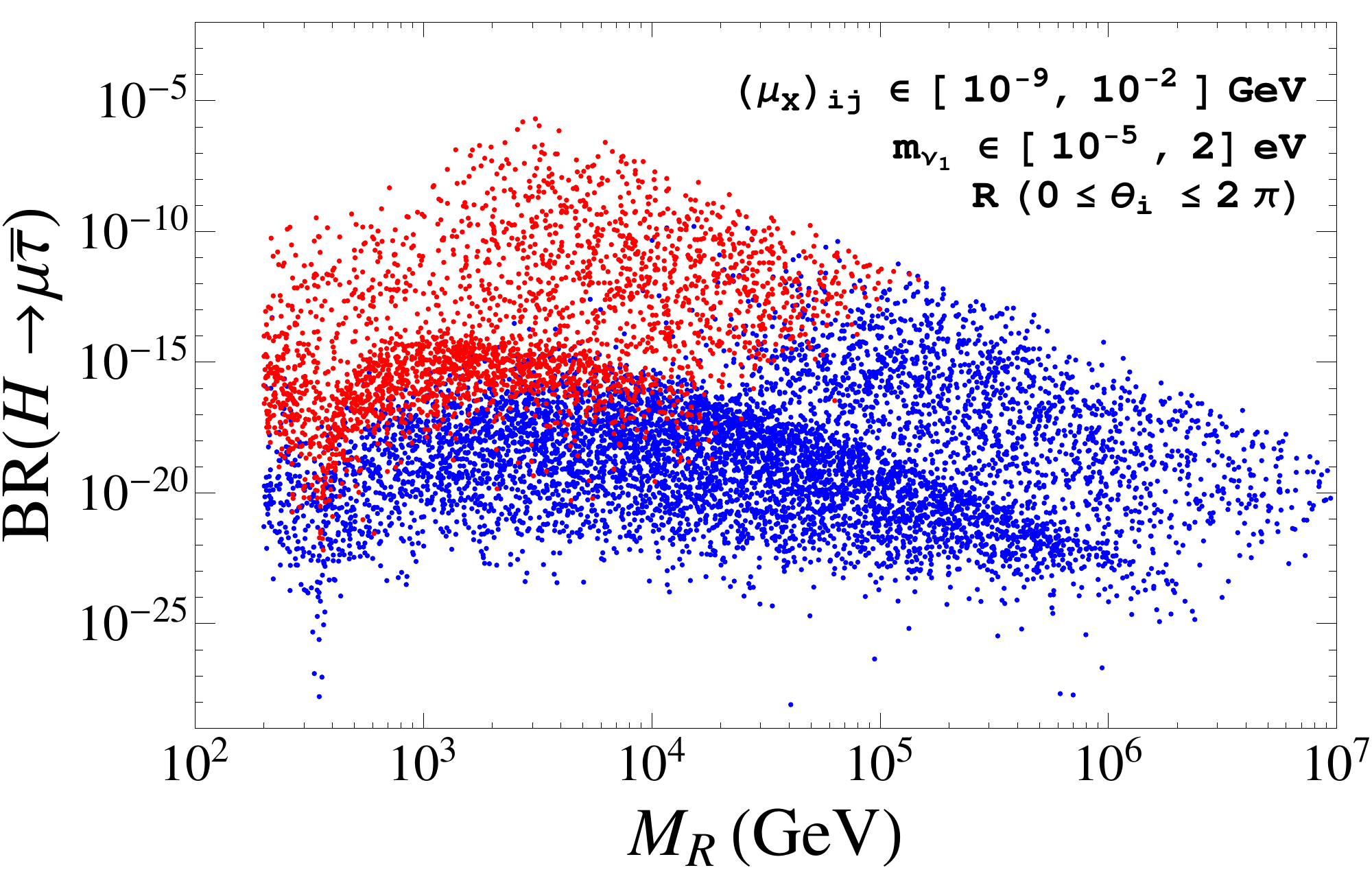}
\end{tabular}
\caption{(left) Contour lines of $\mathrm{BR}(H\rightarrow \mu \bar \tau)$ as a function of both $M_R$ and $\mu_X$. The pink-shaded area is excluded by the upper limit on $\mathrm{BR}(\mu \rightarrow e \gamma)$ from MEG, while the blue-shaded area is excluded by the non-perturbativity of the neutrino Yukawa couplings. (right) Scatter plot for $\mathrm{BR}(H\rightarrow \mu \bar \tau)$ as a function of $M_R$. Red points are excluded by the MEG limit on $\mathrm{BR}(\mu \rightarrow e \gamma)$ while blue points are allowed by all the constraints. }
\label{ContourPlot_degenerate}
\end{center}
\end{figure}

In a degenerate scenario, once the light neutrino masses and mixing are fixed, the only remaining free parameters are $M_R$ and $\mu_X$. This allows us to search for the largest ${\rm BR}(H\to\mu\bar\tau)$ by the means of a contour-line plot like Fig.~\ref{ContourPlot_degenerate}. From Fig.~\ref{BRs_MR_degenerate}, we expect the largest branching ratio to be found at large $M_R$ and small $\mu_X$, where the limits from the neutrino Yukawa couplings perturbativity and the upper bound on $\mathrm{BR}(\mu \rightarrow e \gamma)$ intersect. Looking at Fig.~\ref{ContourPlot_degenerate},
we found that this corresponds to ${\rm BR}(H\to\mu\bar\tau)~\sim 10^{-10}$, which is found for $M_R\sim 2\times 10^4\,{\rm GeV}$ and $\mu_X \sim  5\times 10^{-8} \, {\rm GeV}$. To conclude with more generality, we have randomly scanned over the ISS parameter space in a degenerate scenario.
As can be seen in the right hand plot of Fig.~\ref{ContourPlot_degenerate}, the maximum allowed branching ratio is ${\rm BR}(H\to\mu\bar\tau)\sim 10^{-10}$, in agreement with the result derived from the contour-line plot.

\begin{figure}[t!]
\begin{center}
\begin{tabular}{cc}
\includegraphics[width=.474\textwidth]{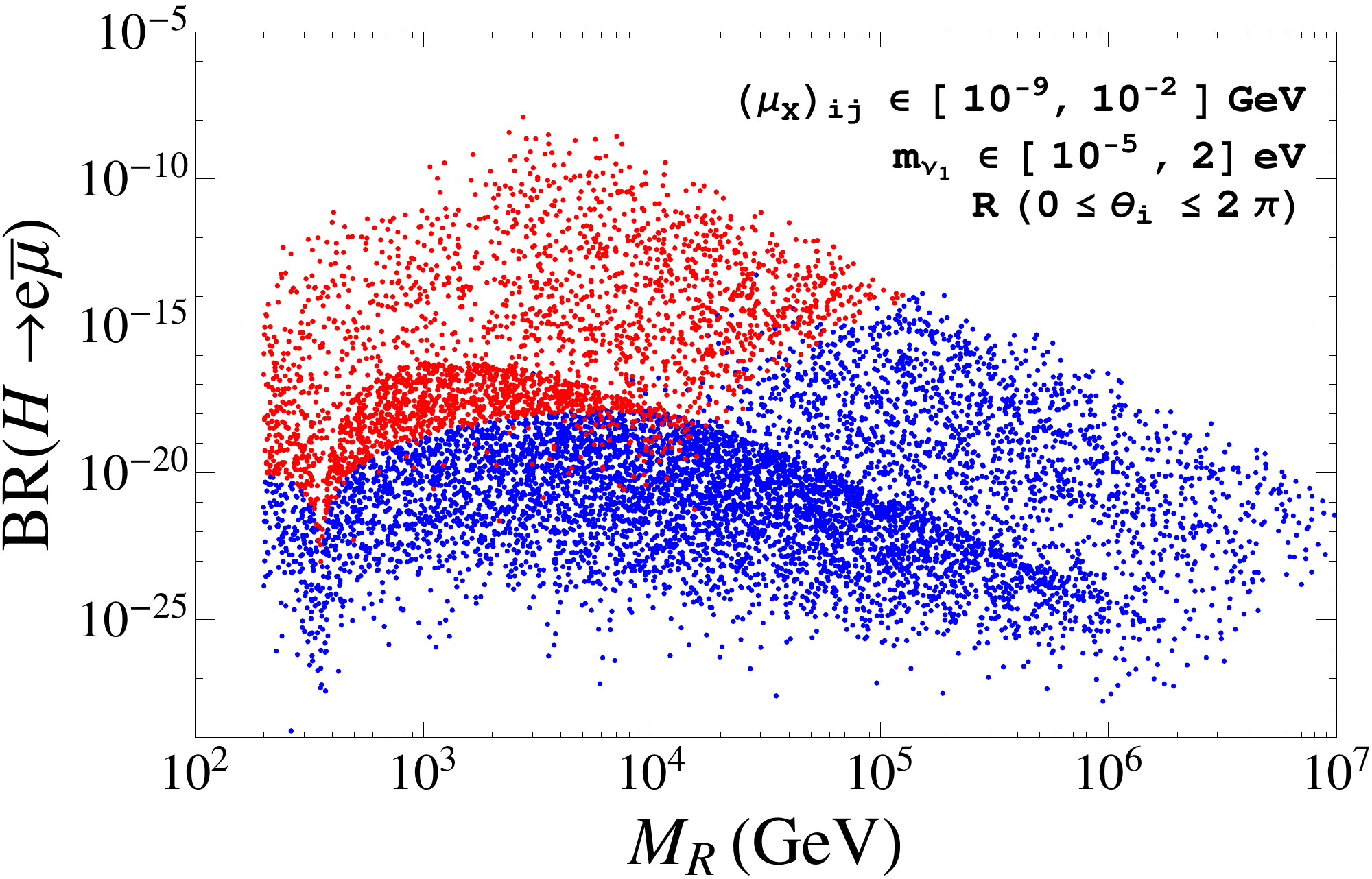} &
\includegraphics[width=.474\textwidth]{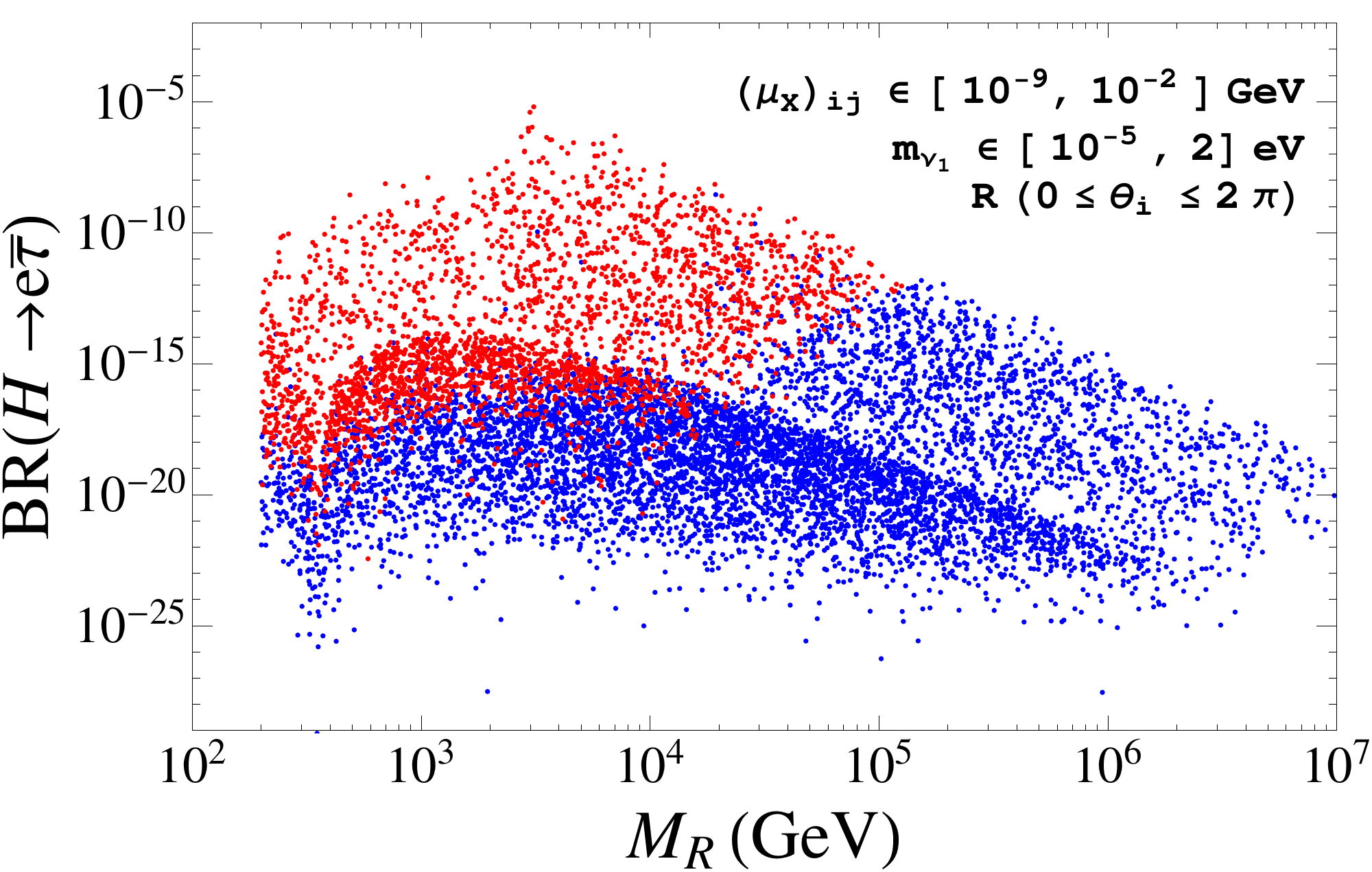}
\end{tabular}
\caption{Scatter plots for $\mathrm{BR}(H\rightarrow e \bar \mu)$ (left) and $\mathrm{BR}(H\rightarrow e \bar \tau)$ (right) as functions of $M_R$. Red points are excluded by the MEG limit on $\mathrm{BR}(\mu \rightarrow e \gamma)$ while blue points are allowed by all the constraints.}
\label{scatterplots}
\end{center}
\end{figure}

We have focussed here on the decay $H\to\mu\bar\tau$ but the other cLFV Higgs decays exhibit the same properties. Using similar random scans on the ISS parameter space, we obtained the plots of Fig.~\ref{scatterplots}
which can be used to conclude that, in a degenerate scenario, the other decays have maximal branching ratios of $\mathrm{BR}(H \to e \bar \tau) \sim 10^{-10}$ and
$\mathrm{BR}(H \to e \bar \mu) \sim 10^{-13}$. We have also performed this study in the hierarchical scenario finding that cLFV Higgs decay rates can be enhanced by as much as one order of magnitude with respect to the degenerate scenario. More details and the corresponding plots can be found in our main article~\cite{Arganda:2014dta}.

\section{Conclusion}

In this work, we have studied cLFV Higgs decays in the inverse seesaw model, where the SM is extended by three pairs of fermionic singlets. Using a full one-loop calculation of the partial decay width for $H\rightarrow \mu \bar \tau,e\bar\tau,e\bar\mu$, we have carefully studied the dependence on the parameters of the model, finding that the main constraints are the upper limit on $\mathrm{BR}(\mu \rightarrow e \gamma)$ and the perturbativity of the neutrino Yukawa couplings. Taking them into account, we conclude that the maximal allowed cLFV Higgs decay rates are  for $H \to e \bar \tau$ and $H \to \mu \bar \tau$, reaching at most $\mathrm{BR}\sim10^{-10}$ for the degenerate heavy neutrinos case and $\mathrm{BR}\sim10^{-9}$ for the hierarchical case. While LHC experiments will not be sensitive to branching ratios so small, this should not deter the searches for cLFV Higgs decays since they are a powerful probe that would help to discriminate between extensions of the Standard Model.

\section*{Acknowledgements}

C.~W. wishes to thank the Moriond organizing committee for its financial support that allowed him to attend the conference. This work is supported by the European Union FP7 ITN INVISIBLES (Marie Curie Actions, PITN-GA-2011-289442), by the CICYT through the project FPA2012-31880,  
by the Spanish Consolider-Ingenio 2010 Programme CPAN (CSD2007-00042) and by the Spanish MINECO's ``Centro de Excelencia Severo Ochoa'' Programme under grant SEV-2012-0249.
E.~A. is financially supported by the Spanish DGIID-DGA grant 2013-E24/2 and the Spanish MICINN grants FPA2012-35453 and CPAN-CSD2007-00042.
X.~M. is supported through the FPU grant AP-2012-6708.

\end{document}